\documentclass[aps,pra,10pt,superscriptaddress,twocolumn,hidelinks,longbibliography,floatfix]{revtex4-2}
\usepackage{graphicx}
\usepackage{amsmath,amsthm,amssymb,dsfont,adjustbox}
\usepackage[normalem]{ulem}
\usepackage{mdframed}
\usepackage{orcidlink}
\newcommand{\ket}[1]{\left| #1 \right\rangle}

\usepackage{hyperref,cleveref}
\usepackage{tikz}
\usetikzlibrary{calc,qcircuit,quotes,quantikz2}

\usepackage[authormarkuptext=name,commentmarkup=uwave]{changes}
\definechangesauthor[name={J\AA L}, color=red]{ja}
\definechangesauthor[name={JRH}, color=blue!70]{jonte}
\newcommand{\jonte}[1]{{\color{blue}[[#1]]$^{\text{JRH}}$}}
\newcommand{\jal}[1]{{\color{green!60!black}[[#1]]$^{\text{J\AA L}}$}}
\definechangesauthor[name={JK}, color=green!70]{jk}
\newcommand{\commentout}[1]{}

\begin{document}

\title{Noncontextual versus contextual interferometry}

\author{Jonte R. Hance\,\orcidlink{0000-0001-8587-7618}}
\email{jonte.hance@newcastle.ac.uk}
\affiliation{School of Computing, Newcastle University, 1 Science Square, Newcastle upon Tyne, NE4 5TG, UK}
\affiliation{Quantum Engineering Technology Laboratories, Department of Electrical and Electronic Engineering, University of Bristol, Woodland Road, Bristol, BS8 1US, UK}

\author{Jakov Krnic}
\affiliation{Department of Electrical Engineering, Linköping University, 581 83 Linköping, Sweden}

\author{Jan-\AA ke Larsson\,\orcidlink{0000-0002-1082-8325}}
\email{jan-ake.larsson@liu.se}
\affiliation{Department of Electrical Engineering, Linköping University, 581 83 Linköping, Sweden}

\begin{abstract}
Feynman famously said that single-particle interference is ``a phenomenon which is impossible to explain in any classical way, and which has in it the heart of quantum mechanics.'' In this paper we show that some of the phenomenology of interference can be reproduced in a ``classical'' way, by reproducing the Elitzur-Vaidman Bomb Tester (including their improved version) using an extension of the quantum simulation logic (QSL) formalism. Our result improves and simplifies a previous result by Catani \emph{et al}, which relies on a much more complicated extension involving a ``toy field theory.'' We also show that not all single-particle interference can be explained by such a simple extension (including that of Catani et al), by showing that Hofmann's three-path interferometer is ``nonclassical'' in a very specific sense: it violates a Kochen-Specker-noncontextual inequality. Given that both our extension of QSL and Catani et al's extension are \emph{noncontextual} --- so do not reproduce the contextual behaviour of Hofmann's three-path interferometer --- the behaviour of that interferometer is a proper example of a phenomenon that has in it the heart of quantum mechanics, according to Feynman. 
\end{abstract}

\maketitle

\section{Introduction}
The Feynman quote about single-particle interference being ``a phenomenon which is impossible to explain in any classical way, and which has in it the heart of quantum mechanics'' \cite{Feynman1965Lect3} is well-known, but in the foundations community it is also well known that this adage is not completely true---we can explain at least some aspects of single-particle interference in a classical way. Recently, Catani \textit{et al} \cite{Catani2023interference} used a ``toy field model'' to do just that. Their model is both Kochen-Specker noncontextual \cite{Kochen1968} and Spekkens' noncontextual \cite{spekkens2005contextuality}. While they do show that some aspects of the phenomenology of single-particle interference can be reproduced using their toy field model, the arguments employed for this are very complex (and feature a number of caveats \cite{Catani2023InterNonclassical,Hance2022CommentCatani}). 
Alongside this, there has been some discussion about the usefulness of the ontological models framework they use \cite{Oldofredi2020Classification,Hance2022Simultaneously,carcassi2024noGoOntic,tezzin2025det,tezzin2025ontological}, and whether representability using a Spekkens' noncontextual ontological model is really both a necessary and sufficient condition for ``classicality'' \cite{budroni2022contextualityReview}.

In this paper, we revisit the issue, aiming to both simplify and clarify the situation. Rather than using the mentioned, rather complicated, toy field model, we show that certain aspects of the phenomenology of single-particle self-interference can be reproduced using a much simpler extension of the toy model, that is straightforward to write down in the notation of Quantum Simulation Logic (QSL) \cite{Johansson2017ClassSim,Johansson2019}. 
These aspects include the phenomenology of the Elitzur-Vaidman Bomb Tester, including its improved versions \cite{Elitzur1993Bomb}.
It follows that the aspects of the phenomenology of interference responsible for the Elitzur-Vaidman bomb tester are indeed explainable ``in a classical way,'' alluding to the Feynman quote. 

We then instead look at more complex aspects of the phenomenology of single-particle interference---using Hofmann's three-path interferometer \cite{Hofmann2023Sequential,Hance2024CounterfactualNegativity,hofmann2025quantum,sagawa2025quantum}---and show that it requires a Kochen-Specker contextual model \cite{Kochen1968,budroni2022contextualityReview}---specifically, that the interferometer can be used to violate the KCBS inequality~\cite{Klyachko2008KCBS}. This shows that Kochen-Specker contextuality is a crucial indicator that a given aspect of the phenomenology of single-particle self-interference cannot be reproduced classically.

As well as providing a clear delineation between which aspects of single-particle self-interference are and are not reproducible classically, this work reinforces the idea we should use Kochen-Specker contextuality as an indicator of a potential quantum resource \cite{amaral2019resource}, which can be used to perform tasks more effectively than a classical solution could, so would constitute a quantum advantage --- as has been shown for specific tasks in e.g., Refs~\cite{Pironio2010BellCertRandomNos,Raussendorf2013ContextualityMBQC,Um2013ContextualRN,Howard2014,BermejoVega2017ContextualityQubitQC,Frembs2018ContextualResourceMBQCQudits,Bharti2019NoncontextualSelftesting,Saha2019ContextualityOneWay,ArvidssonShukur2020Postsel,Gupta2023ContextualityQCommComplex}. Given the current focus on identifying such a quantum advantage, to ensure the current worldwide programme to develop quantum technologies bears useful dividends, this work will be of use not just to the foundations community, but also more widely to the quantum computing/algorithms, quantum sensing, and quantum metrology communities.

\section{Elitzur-Vaidman Bomb Tester in Quantum Simulation Logic}\label{sect.EV}

\begin{figure}
    \centering
    (a) \begin{tikzpicture}[every edge/.style={draw,densely dotted}]
    \scriptsize
    \draw [ultra thick] (1,-.5)--++(.5,0)node[scale=.8,above right]{50:50};
    \draw [ultra thick] (4,-.5)--++(.5,0)node[scale=.8,above right]{50:50};
    \draw (0,-1) edge ["0" above right,at start]+(1,0)++(1,0)edge+(.25,.5)++(.25,.5)
    --++(.25,-.5)to["$\tfrac12$",near start]++(2.5,0)--++(.25,.5)--++(.25,-.5)to["1" above left,at end]++(1,0);
    \draw (0,0) to["1" above right,at start]++(1,0)--++(.25,-.5)--++(.25,.5)
    to["$\tfrac12$",near start]++(2.5,0)--++(.25,-.5)edge+(.25,.5)++(.25,.5)edge["0" above left,at end]++(1,0);
    \draw[red,densely dashed,thick] ({3+0.25*cos(70)},{0+0.25*sin(70)}) arc (70:380:0.25)--({3.06+0.25*cos(20)},{0.06+0.25*sin(20)})--({3.06+0.25*cos(70)},{0.06+0.25*sin(70)})--cycle;
    \end{tikzpicture}\vspace{1em}
    
    (b) \begin{tikzpicture}[every edge/.style={draw,densely dotted}]
    \scriptsize
    \draw [ultra thick] (1,-.5)--++(.5,0)node[scale=.8,above right]{50:50};
    \draw [ultra thick] (4,-.5)--++(.5,0)node[scale=.8,above right]{50:50};
    \draw (0,-1) edge ["0" above right,at start]+(1,0)++(1,0)edge+(.25,.5)++(.25,.5)
    --++(.25,-.5)to["$\tfrac12$",near start]++(2.5,0)--++(.25,.5)--++(.25,-.5)to["$\tfrac14$" above left,at end]++(1,0);
    \draw (0,0) to["1" above right,at start]++(1,0)--++(.25,-.5)--++(.25,.5)
    edge["0",pos=0.85]+(2.5,0)to["$\tfrac12$",pos=.42]+(1.5,0) ++(2.5,0)edge+(.25,-.5)++(.25,-.5)--++(.25,.5)to["$\tfrac14$" above left,at end]++(1,0);
    \fill[red,densely dotted,thick] ({3+0.25*cos(70)},{0+0.25*sin(70)}) arc (70:380:0.25)--({3.06+0.25*cos(20)},{0.06+0.25*sin(20)})--({3.06+0.25*cos(70)},{0.06+0.25*sin(70)})--cycle;
    \end{tikzpicture}
    \caption{The Elitzur-Vaidman Bomb Tester. (a) Bomb doesn't work so it does not interact with the photon. (b) Bomb works, so detects photons that arrive at the bomb. We show in Section~\ref{sect.EV} that we can reproduce the peculiar behaviour of this scenario classically, using a minimal extension of QSL.}
    \label{fig:EV}
\end{figure}
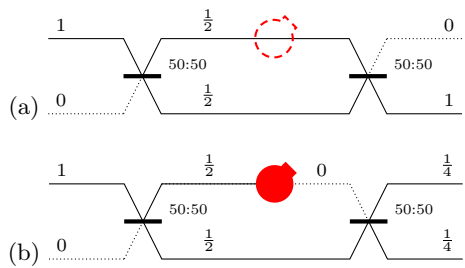

In this first section, we show how that the Elitzur-Vaidman Bomb Tester (Fig.~\ref{fig:EV}) can be reproduced using Quantum Simulation Logic \cite{Johansson2019}, a notational language used to represent and extend Spekkens' Toy Model \cite{Spekkens2007ToyModel}.

Note that, if we represent our interferometer in a second-quantised way, we can assign the state $\ket{00}$ if no photon is present in either beam, $\ket{01}$ if one photon present in the lower beam and none in the upper, and $\ket{10}$ if one photon is present in the upper beam and none in the lower. (This is a different choice than the states $\ket{1}$ and $i\ket{2}$ used by Elitzur and Vaidman \cite{Elitzur1993Bomb}. Note that we avoid an extra phase on $i\ket{2}$, which removes the need for complex coefficients below.) This gives the beamsplitter map
\begin{equation}
\begin{split}
\ket{00}&\mapsto\ket{00},\\
\ket{01}&\mapsto\sqrt{\tfrac12}\ket{01}-\sqrt{\tfrac12}\ket{10},\text{ and}\\
\ket{10}&\mapsto\sqrt{\tfrac12}\ket{01}+\sqrt{\tfrac12}\ket{10},\\
\end{split}
\end{equation}
Since we restrict ourselves to the zero- and one-photon subspace of Fock space we can choose $\ket{11}\mapsto\ket{11}$ to generate a unitary that acts on the qubit subspace and preserves photon number in our construction. 
This does not reproduce the true beamsplitter map in the two-photon subspace (or higher photon number spaces), but will serve our purposes here given we restrict ourselves to the zero- and one-photon subspaces.
It is easy to check that a quantum circuit for this map can be obtained by first applying a $CNOT$ (a Pauli-$X$ controlled by qubit 1), then a $R_y(\pi/2)$ rotation controlled by qubit 2 onto qubit 1, then another $CNOT$ (controlled by qubit~1). 

In the Quantum Simulation Logic \cite{Johansson2019} notation for Spekkens' toy model \cite{Spekkens2007ToyModel}, the state of the two second-quantised qubits is represented by two pairs of bits $(z_1,x_1)(z_2,x_2)$, and the CNOT map is 
\begin{equation}
(z_1,x_1)(z_2,x_2)\xrightarrow{CNOT}(z_1,x_1+x_2)(z_1+z_2,x_2),
\end{equation}
(here and below addition is modulo 2, and $\overline x=x+1$).
A~QSL $R_y(\pi/2)$ gate performs
\begin{equation}
(z,x)\xrightarrow{R_y(\pi/2)}(x,\overline z).
\end{equation}
A\ controlled-$R_y(\pi/2)$, or $CR_y(\pi/2)$, gate controlled by qubit 2 is not Clifford, and is not present in either QSL or the toy model, but we can construct a ``kickback-free'' version by ignoring the phase kickback of the control, as
\begin{equation}
\begin{split}
(z_1,&x_1)(z_2,x_2)\xrightarrow{C_{21}R_y(\pi/2)}
\\&\bigr(x_1z_2+z_1\overline{z_2},\overline z_1z_2+x_1\overline{z_2}\bigr)(z_2,x_2),
\end{split}
\end{equation}
so that the $R_y(\pi/2)$ gate is performed only if the computational bit for system 2 is set to 1.
\begin{figure}
\centering
\begin{tikzpicture}
\node (a) {$\sim$};
\draw (a)+(-.5,0) node[anchor=east] {\includegraphics{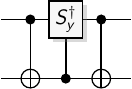}};
\draw (a)+(.5,0) node[anchor=west] {\includegraphics{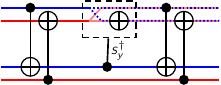}};
\end{tikzpicture}
\caption{Zero- or one-photon beam-splitter in second quantization (left), Quantum Simulation Logic circuit that reproduces noncontextual behaviour mimicking this beamsplitter (right).}
\label{fig:beamsplitter}
\end{figure}
The whole map can now be viewed in Fig.~\ref{fig:beamsplitter}.
The initial QSL state for a single photon in a definite path is $(z,x)(\overline z,y)$ where $z=1$ if a photon is sent into the upper input path, or $z=0$ into the lower path, and $x$ and $y$ are fair independent coin tosses. 
The first beamsplitter map is 
\begin{equation}
    \begin{split}
        &(z,x)(\overline z,y)
        \xrightarrow{CNOT}(z,x+y)(1,y)\\
        &\quad\xrightarrow{C_{21}R_y(\pi/2)}(x+y,\overline z)(1,y)\\
        &\quad\xrightarrow{CNOT}(x+y,y+\overline z)(\overline {x+y},y),
    \end{split}
\end{equation}
where there now is equal probability 1/2 for a photon either in the upper output path ($x+y=1$) or in the lower output path ($x+y=0$).

Adding a second beamsplitter to form an interferometer will repeat the process, so that
\begin{equation}
    \begin{split}
        (x+y,y+\overline z)&(\overline {x+y},y)
        \xrightarrow{CNOT}(x+y,\overline z)(1,y)\\
        &\xrightarrow{C_{21}R_y(\pi/2)}(\overline z,\overline{x+y})(1,y)\\
        &\xrightarrow{CNOT}(\overline z,\overline x)(z,y).
    \end{split}
\end{equation}
A photon entering the upper input path of such an interferometer ($z=1$) will exit in the lower output path, and vice versa.

With a bomb to be tested in the upper internal path of the interferometer we now want to consider a photon entering the upper path ($z=1$) in two possible scenarios --- one where the bomb does not work, and one where the bomb does work.
In the first case, the interferometer behaves as just described, but in the second the bomb removes any photon present on the upper internal path in the interferometer. 
Indeed, it performs a measurement that randomises the phase of the upper path, we here add an extra random bit~$w$ to make the total map
\begin{equation}\label{Eq.7}
    \begin{split}
        (x+y,y)(\overline {x+y},y)&\xrightarrow{\text{Bomb}}(0,w)(\overline {x+y},y).
    \end{split}
\end{equation}
With probability 1/2, the bomb will have exploded ($x+y=1$) and then there is no photon present, in which case the second beamsplitter will leave the state unchanged. 
Otherwise, the second beamsplitter will perform
\begin{equation}\label{Eq.8}
    \begin{split}
        (0,w)&(1,y)
        \xrightarrow{CNOT}(0,w+y)(1,y)\\
        &\xrightarrow{C_{21}R_y(\pi/2)}
        (w+y,1)(1,y)\\
        &\xrightarrow{CNOT}(w+y,\overline y)(\overline {w+y},y)
    \end{split}
\end{equation}
After the second beamsplitter, there is equal probability 1/4 of detection either in the upper path ($w+y=1$) or the lower path ($w+y=0$). 

We can compare this with the case where the bomb works but just randomises the phase in the upper path, essentially performing a non-demolition measurement, and leaving the photon intact.
In this case, the second beamsplitter will perform
\begin{equation}\label{Eq.9}
    \begin{split}
        (x+y,w)&(\overline {x+y},y)
        \xrightarrow{CNOT}(x+y,w+y)(1,y)\\
        &\xrightarrow{C_{21}R_y(\pi/2)}(w+y,\overline {x+y})(1,y)\\
        &\xrightarrow{CNOT}(w+y,\overline x)(\overline {w+y},y)
    \end{split}
\end{equation}
Here, after the second beamsplitter there is equal probability 1/2 of detection either in the upper path ($w+y=1$) or in the lower path ($w+y=0$). 

In both cases the behaviour of the second beamsplitter is what enables reproducing the bomb tester phenomenology.
The first $CNOT$ stores the difference of the computational coordinates (in quantum mechanics, the computational-basis ket indices) in the second system; it checks if the computational coordinates are equal.
If the incoming state is in the single-photon subspace this comparison outcome is nonzero, and only then the $R_y(\pi/2)$ rotation is performed.
The second $CNOT$ then uncomputes the comparison outcome present in system 2.
It is the phase kickback of the two $CNOT$s that enables the behaviour.
Through this mechanism we can reproduce the phenomenology of the standard (50:50) Elitzur-Vaidman bomb tester with minimal extensions of QSL, or equivalently (here) the toy model, well-established models for being able to represent a phenomenon classically.

\section{The Elitzur-Vaidman Improved Bomb Tester with Quantum Simulation Logic}\label{sect.EVImproved}

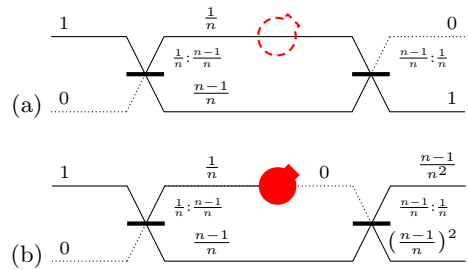
\begin{figure}
    \centering
    (a) \begin{tikzpicture}[every edge/.style={draw,densely dotted}]
    \scriptsize
    \draw [ultra thick] (1,-.5)--++(.5,0)node[scale=.8,above right]{$\tfrac1n{:}\tfrac{n-1}n$};
    \draw [ultra thick] (4,-.5)--++(.5,0)node[scale=.8,above right]{$\tfrac{n-1}n{:}\tfrac1n$};
    \draw (0,-1) edge ["0" above right,at start]+(1,0)++(1,0)edge+(.25,.5)++(.25,.5)
    --++(.25,-.5)to["$\tfrac{n-1}n$",near start]++(2.5,0)--++(.25,.5)--++(.25,-.5)to["1" above left,at end]++(1,0);
    \draw (0,0) to["1" above right,at start]++(1,0)--++(.25,-.5)--++(.25,.5)
    to["$\tfrac1n$",near start]++(2.5,0)--++(.25,-.5)edge+(.25,.5)++(.25,.5)edge["0" above left,at end]++(1,0);
    \draw[red,densely dashed,thick] ({3+0.25*cos(70)},{0+0.25*sin(70)}) arc (70:380:0.25)--({3.06+0.25*cos(20)},{0.06+0.25*sin(20)})--({3.06+0.25*cos(70)},{0.06+0.25*sin(70)})--cycle;
    \end{tikzpicture}\vspace{1em}
    
    (b) \begin{tikzpicture}[every edge/.style={draw,densely dotted}]
    \scriptsize
    \draw [ultra thick] (1,-.5)--++(.5,0)node[scale=.8,above right]{$\tfrac1n{:}\tfrac{n-1}n$};
    \draw [ultra thick] (4,-.5)--++(.5,0)node[scale=.8,above right]{$\tfrac{n-1}n{:}\tfrac1n$};
    \draw (0,-1) edge ["0" above right,at start]+(1,0)++(1,0)edge+(.25,.5)++(.25,.5)
    --++(.25,-.5)to["$\tfrac{n-1}n$",near start]++(2.5,0)--++(.25,.5)--++(.25,-.5)to["$\bigl(\tfrac{n-1}{n}\bigr)^2$" above left,at end]++(1,0);
    \draw (0,0) to["1" above right,at start]++(1,0)--++(.25,-.5)--++(.25,.5)
    edge["0",pos=0.85]+(2.5,0)to["$\tfrac1n$",pos=.42]+(1.5,0) ++(2.5,0)edge+(.25,-.5)++(.25,-.5)--++(.25,.5)to["$\tfrac{n-1}{n^2}$" above left,at end]++(1,0);
    \fill[red,densely dotted,thick] ({3+0.25*cos(70)},{0+0.25*sin(70)}) arc (70:380:0.25)--({3.06+0.25*cos(20)},{0.06+0.25*sin(20)})--({3.06+0.25*cos(70)},{0.06+0.25*sin(70)})--cycle;
    \end{tikzpicture}
    \caption{The Elitzur-Vaidman improved bomb tester that uses unbalanced beamsplitters. (a) Bomb doesn't work so it does not interact with the photon. (b) Bomb works, so detects photons that arrive at the bomb. We show in Section~\ref{sect.EVImproved} that we can reproduce the peculiar behaviour of this scenario classically, using a small extension of QSL.}
    \label{fig:EV2}
\end{figure}

We now extend QSL further to be able to reproduce the behaviour of the Elitzur-Vaidman improved Bomb Tester (\cite{Elitzur1993Bomb}, see Fig.~\ref{fig:EV2}).
This uses two unbalanced beamsplitters - the quantum map for these beamsplitters, with the above-used restriction to the zero- or one-photon subspace, would now be
\begin{equation}
\begin{split}
\ket{00}&\mapsto\ket{00},\\
\ket{01}&\mapsto\cos\theta\ket{01}-\sin\theta\ket{10},\text{ and}\\
\ket{10}&\mapsto\sin\theta\ket{01}+\cos\theta\ket{10},\\
\end{split}
\end{equation}
with $\theta=\arccos\sqrt{1/n}$ for the first beamsplitter and $\theta=\arccos\sqrt{(n-1)/n}$ for the second, if $n=2$ both reduce to the simple symmetric 50:50 case. 
Also here a quantum circuit that performs the first beamsplitter map can be obtained by first applying a $CNOT$ (a Pauli-$X$ controlled by qubit 1), then a $R_y(\theta)$ rotation controlled by qubit 2 onto qubit 1, then another $CNOT$ (controlled by qubit 1), and similar for the second. 

Quantum Simulation Logic does not have enough resolution in the Bloch sphere $XZ$ meridian to model this, but can be extended to rational-ratio $k/n:(n-k)/n$ beamsplitters as follows. 
The state of the two second-quantised qubits is then two tuples $(z_1,y_1)(z_2,y_2)$, where $z_i$ are bits and $y_i$ are integers mod $n$.
For $n=2$ the below description is equivalent to the extended QSL of Section~\ref{sect.EV}, but involves the coordinate change $y_i=x_i+z_i$ mod 2.

Also here the first $CNOT$ is used to compare the computational coordinates, to perform the rotation if the difference is nonzero.
Remembering that addition and subtraction are equivalent mod 2 but not mod $n$, we must subtract rather than add the now higher-resolution phase coordinates from/to each other in the first $CNOT$, and perform addition only as uncomputation in the second $CNOT$.
We therefore need two variants of the $CNOT$,
\begin{equation}
  (z_1,y_1)(z_2,y_2)\xrightarrow{CNOT_\pm}(z_1,y_1\pm y_2)(z_2\pm z_1,y_2),
\end{equation}
where we again note that $z_1+z_2=z_1-z_2$ mod 2. 
We can extend the QSL rotation $R_y(\theta)$ to this case with $\theta=\arccos\sqrt{k/n}$ through
\begin{equation}
(z,x)\xrightarrow{R_y(\theta)}(z+\delta_{xkn}, x - k),
\end{equation}
where $\delta_{xkn}=1$ if there occurs unsigned-integer overflow when adding $n-k$ to the phase coordinate $x$ register mod $n$ ($\equiv$ subtracting $k$ from $x$, more formally when the smallest non-negative representative of $x$ mod $n$ is $\ge k$) and $\delta_{xkn}=0$ otherwise.
We note that $p(\delta_{Xkn}=0)=k/n$ if $p(X=x)=1/n$ for $0\le x\le n-1$; with a random phase the computational coordinate $z$ is unchanged with probability $k/n$.  
A~kickback-free $C_{21}R_y(\theta)$ gate can be obtained as
\begin{equation}
\begin{split}
(z_1,&x_1)(z_2,x_2)\xrightarrow{C_{21}R_y(\theta)}\\
&\bigl(z_1+z_2\delta_{x_1kn},x_1-z_2k\bigr)(z_2,x_2).
\end{split}
\end{equation}

The whole map is still similar to that in Fig.~\ref{fig:beamsplitter}.
The initial QSL state for a single photon in a definite path is $(z,x)(\overline z,y)$ where $z=1$ corresponds to a photon in the upper input path and $z=0$ a photon in the lower path, and $x$ and $y$ are random values mod $n$.
A rational-ratio $k/n:(n-k)/n$ beamsplitter map performs 
\begin{equation}
    \begin{split}
        &(z,x)(\overline z,y)
        \xrightarrow{CNOT_-}(z,x-y)(1,y)\\
        &\xrightarrow{C_{21}R_y(\theta)}(z +\delta_{(x-y)kn},x-y - k)(1,y)\\
        &\xrightarrow{CNOT_+}\bigl(z+\delta_{(x-y)kn},x-k)(\overline z+\delta_{(x-y)kn}, y).
    \end{split}
    \raisetag{3.3em}
\end{equation}
The random phase difference $x-y$ gives a $k/n$ probability that a photon will remain in the same path.

Adding a subsequent $(n-k)/n:k/n$ beamsplitter to form an interferometer will add a second phase coordinate shift, here adding $n-(n-k)=k$.
The two shifts combine so that there either is an overflow in the first beamsplitter or in the second, but never in both, so there will in total be 1 added to $z$ in such an interferometer, so that
\begin{equation}
    \begin{split}
        &( z+\delta_{(x-y)kn},x-k)(\overline z+\delta_{(x-y)kn}, y)\\
        &\xrightarrow{CNOT_-}(z+\delta_{(x-y)kn},x-y-k)(1,y)\\
        &\xrightarrow{C_{21}R_y(\theta)}(\overline z,x-y)(1,y)\\
        &\xrightarrow{CNOT_+}\bigl(\overline z,x)(z, y).
    \end{split}
\end{equation}
A photon entering the upper input path of such an interferometer will exit in the lower output path, and vice versa.

Again, we insert a bomb to be tested in the upper internal path of the interferometer and enter a photon in the upper path ($z=1$).
With a non-working bomb, the interferometer behaves as just described, and with a working bomb, any photon present on the upper internal path is removed and the phase randomized,
\begin{equation}
    \begin{split}
        &(\overline{\delta_{(x-y)kn}},x - k)(\delta_{(x-y)kn}, y)
        \\&\quad
        \xrightarrow{\text{Bomb}}(0,w)(\delta_{(x-y)kn}, y).
    \end{split}
\end{equation}
With probability $k/n$, the bomb will have exploded, and then there is no photon present as before. 
Otherwise, the second beamsplitter will perform
\begin{equation}
    \begin{split}
        (0&, w)(1,y) 
        \xrightarrow{CNOT_-}(0,w-y)(1,y) \\
        &\xrightarrow{C_{21}R_y(\theta)}
        ({\delta_{(w-y)(n-k)n}},w-y+k)(1,y)\\
        &\xrightarrow{CNOT_+}({\delta_{(w-y)(n-k)n}}, w+k)(\overline{\delta_{(w-y)(n-k)n}},y).
    \end{split}
\end{equation}
The random phase difference $w-y$ gives a conditional probability $(n-k)/n$ for the photon remaining in the same (lower) path, so a total probability $(n-k)^2/n^2$.  
The probability to find a photon in the upper path is $k(n-k)/n^2$; these are the same predictions as from QM for the improved EV bomb tester \cite{Elitzur1993Bomb}.

We can compare this with the case where the bomb works but just randomises the phase in the upper path, where the second beamsplitter performs
\begin{equation}
    \begin{split}
        &\bigl(z+\delta_{(x-y)kn},w)(\overline z+\delta_{(x-y)kn}, y)\xrightarrow{CNOT_-}
        \\&(z+\delta_{(x-y)kn},w-y)(1,y)\xrightarrow{C_{21}R_y(\theta)}\\
        &(z+\delta_{(x-y)kn}+\delta_{(w-y)(n-k)n},w-y+k)(1,y)\\
        &\xrightarrow{CNOT_+}\bigl(z+\delta_{(x-y)kn}+\delta_{(w-y)(n-k)n},w+k)\\
        &\qquad\qquad\qquad\quad
        (\overline z+\delta_{(x-y)kn}+\delta_{(w-y)(n-k)n}, y).
    \end{split}
\end{equation}
The two independent phase differences $x-y$ and $w-y$ will give independent random outputs, the first will output~0 with probability $k/n$ and the second with probability $(n-k)/n$.
If a photon is sent into the upper input path, finding a photon in the upper output path would require no change or two changes, which both have probability $k(n-k)/n^2$. The total probability is $2k(n-k)/n^2$, just as predicted by QM. 

That we only have access to rational-ratio beamsplitters is not a restriction; we note that the set of rationals is dense in the set of reals, so we may approximate a irrational-ratio beamsplitter to arbitrary precision.
Thus, we can also reproduce the complete phenomenology of the improved Elitzur-Vaidman bomb tester \cite{Elitzur1993Bomb} with only small extensions to QSL.
While far simpler than Catani et al's analysis in \cite{Catani2023interference}, both Sections \ref{sect.EV} and \ref{sect.EVImproved} show the same result as this --- that there are aspects of the phenomenology of (quantum) interference which are reproducible ``classically''.

\section{Contextual Interferometry}\label{sect:ContextInterferometry}

In this section, we now aim to show aspects of the phenomenology of single-particle interference that are not reproducible ``classically''. 
We note that reproducibility by Spekkens' toy model was used as a criterion for classicality in \cite{Catani2023interference}. However, here we will extend that criterion to say a phenomenon isn't reproducible ``classically'' if it is not reproducibility by any Kochen-Specker noncontextual model \cite{Kochen1968,budroni2022contextualityReview}. Spekkens' toy model is Kochen-Specker-noncontextual, so this extended notion of non-classicality includes the former notion. 

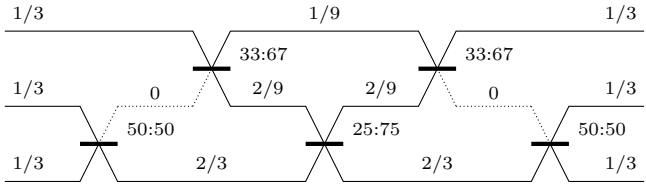
\begin{figure}[ht]
\begin{tikzpicture}[every edge/.style={draw,densely dotted}]
    \scriptsize
    \draw [ultra thick] (1,-.5)--++(.5,0)node[above right]{50:50};
    \draw [ultra thick] (2.5,.5)--++(.5,0)node[above right]{33:67};
    \draw [ultra thick] (4,-.5)--++(.5,0)node[above right]{25:75};
    \draw [ultra thick] (5.5,.5)--++(.5,0)node[above right]{33:67};
    \draw [ultra thick] (7,-.5)--++(.5,0)node[above right]{50:50};
    \draw (0,1) to ["1/3" above right, at start]++(2.5,0)--++(.25,-.5)--++(.25,.5)
    to ["1/9"]++(2.5,0)--++(.25,-.5)--++(.25,.5) to ["1/3" above left,at end]++(2.5,0);
    \draw (0,0) to ["1/3" above right,at start]++(1,0)--++(.25,-.5)
    edge+(.25,.5)++(.25,.5)edge["0"]++(1,0)++(1,0)edge+(.25,.5)++(.25,.5)--++(.25,-.5)
    to ["2/9"]++(1,0)--++(.25,-.5)--++(.25,.5)to["2/9"]++(1,0)--++(.25,.5)
    edge+(.25,-.5)++(.25,-.5)edge["0"]++(1,0)++(1,0)edge+(.25,-.5)++(.25,-.5)
    --++(.25,.5)to["1/3" above left,at end]++(1,0);
    \draw (0,-1) to["1/3" above right,at start]++(1,0)--++(.25,.5)--++(.25,-.5)
    to["2/3"]++(2.5,0)--++(.25,.5)--++(.25,-.5)to["2/3"]++(2.5,0)
    --++(.25,.5)--++(.25,-.5)to["1/3" above left,at end]++(1,0);
    \end{tikzpicture}
\caption{Detection probabilities in a three-path interferometer with indicated beamsplitter ratios. Input state $\frac1{\sqrt3}(\ket{100}+\ket{010}+\ket{001})$.}
\label{fig:interferometer}
\end{figure}

In this situation it is of interest to find an interferometer that requires contextuality. Hofmann's three-path interferometer is a good example, containing (in Hofmann's own words) five contexts---however, his original paper \cite{Hofmann2023Sequential} does not contain an analysis in terms of Kochen-Specker contextuality, so we will provide such an analysis.
Again we use the single-photon subspace of a second-quantised description, using the basis $\ket{001}$ for a photon present in the lowest beam, $\ket{010}$ for a photon present in the middle beam, and $\ket{100}$ for a photon present in the top beam.
The state evolution through the interferometer of our chosen input state is, with our phase conventions,
\begin{equation}
\begin{split}
\sqrt{\tfrac13}&\ket{100}+\sqrt{\tfrac13}\ket{010}+\sqrt{\tfrac13}\ket{001}\\
&\xrightarrow{BS_1}\sqrt{\tfrac13}\ket{100}+\sqrt{\tfrac23}\ket{001}\\
&\xrightarrow{BS_2}\sqrt{\tfrac19}\ket{100}+\sqrt{\tfrac29}\ket{010}+\sqrt{\tfrac23}\ket{001}\\
&\xrightarrow{BS_3}\sqrt{\tfrac19}\ket{100}-\sqrt{\tfrac29}\ket{010}+\sqrt{\tfrac23}\ket{001}\\
&\xrightarrow{BS_4}\sqrt{\tfrac13}\ket{100}+\sqrt{\tfrac23}\ket{001}\\
&\xrightarrow{BS_5}\sqrt{\tfrac13}\ket{100}-\sqrt{\tfrac13}\ket{010}+\sqrt{\tfrac13}\ket{001}\\
\label{eq:contextual_state}
\end{split}
\end{equation}

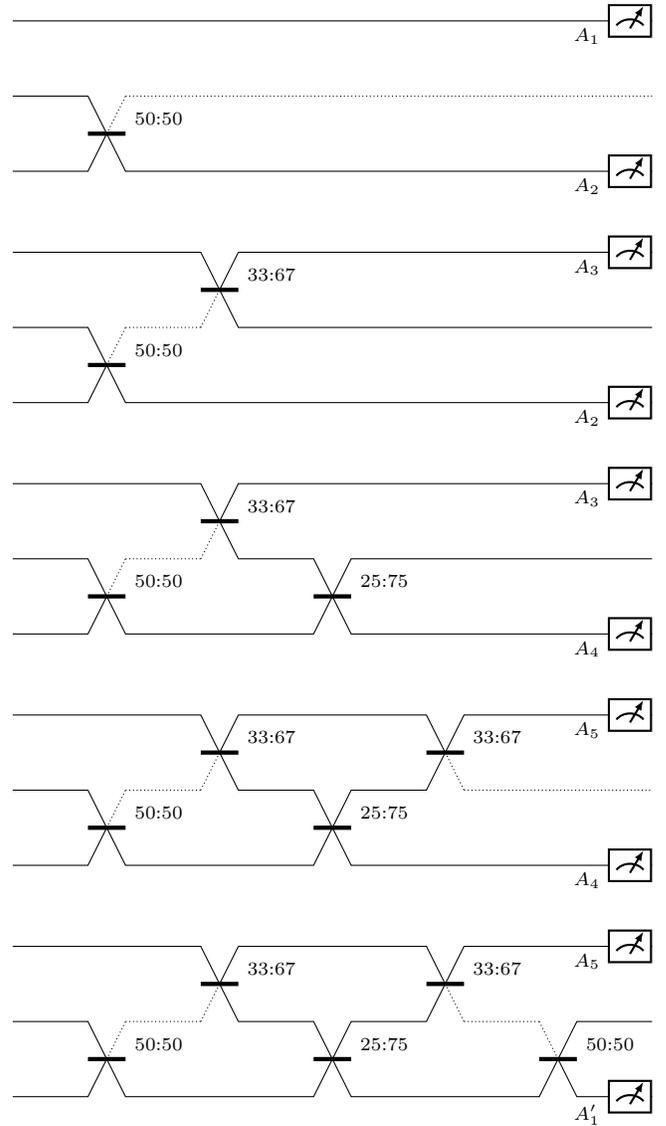
\begin{figure}
\begin{tikzpicture}[every edge/.style={draw,densely dotted}]
    \scriptsize
    \draw [ultra thick] (1,-.5)--++(.5,0)node[above right]{50:50};
    \draw (0,1) --++(8.5,0)
    node[meter,anchor=east](m){}(m.west)node[below left]{$A_1$};
    \draw (0,0) --++(1,0)--++(.25,-.5)edge+(.25,.5)++(.25,.5)edge++(7,0);
    \draw (0,-1) --++(1,0)--++(.25,.5)--++(.25,-.5)--++(7,0)
    node[meter,anchor=east](m){}(m.west)node[below left]{$A_2$};
\end{tikzpicture}\bigskip

\begin{tikzpicture}[every edge/.style={draw,densely dotted}]
    \scriptsize
    \draw [ultra thick] (1,-.5)--++(.5,0)node[above right]{50:50};
    \draw [ultra thick] (2.5,.5)--++(.5,0)node[above right]{33:67};
    \draw (0,1) --++(2.5,0)--++(.25,-.5)--++(.25,.5)--++(5.5,0)
    node[meter,anchor=east](m){}(m.west)node[below left]{$A_3$};
    \draw (0,0) --++(1,0)--++(.25,-.5)
    edge+(.25,.5)++(.25,.5)edge+(1,0)++(1,0)edge+(.25,.5)++(.25,.5)
    --++(.25,-.5)--++(5.5,0);
    \draw (0,-1) --++(1,0)--++(.25,.5)--++(.25,-.5)--++(7,0)
    node[meter,anchor=east](m){}(m.west)node[below left]{$A_2$};
\end{tikzpicture}\bigskip

\begin{tikzpicture}[every edge/.style={draw,densely dotted}]
    \scriptsize
    \draw [ultra thick] (1,-.5)--++(.5,0)node[above right]{50:50};
    \draw [ultra thick] (2.5,.5)--++(.5,0)node[above right]{33:67};
    \draw [ultra thick] (4,-.5)--++(.5,0)node[above right]{25:75};
    \draw (0,1) --++(2.5,0)--++(.25,-.5)--++(.25,.5)--++(5.5,0)
    node[meter,anchor=east](m){}(m.west)node[below left]{$A_3$};
    \draw (0,0) --++(1,0)--++(.25,-.5)
    edge+(.25,.5)++(.25,.5)edge+(1,0)++(1,0)edge+(.25,.5)++(.25,.5)
    --++(.25,-.5)--++(1,0)--++(.25,-.5)--++(.25,.5)--++(4,0);
    \draw (0,-1) --++(1,0)--++(.25,.5)--++(.25,-.5)--++(2.5,0)--++(.25,.5)--++(.25,-.5)--++(4,0)
    node[meter,anchor=east](m){}(m.west)node[below left]{$A_4$};
\end{tikzpicture}\bigskip

\begin{tikzpicture}[every edge/.style={draw,densely dotted}]
    \scriptsize
    \draw [ultra thick] (1,-.5)--++(.5,0)node[above right]{50:50};
    \draw [ultra thick] (2.5,.5)--++(.5,0)node[above right]{33:67};
    \draw [ultra thick] (4,-.5)--++(.5,0)node[above right]{25:75};
    \draw [ultra thick] (5.5,.5)--++(.5,0)node[above right]{33:67};
    \draw (0,1) --++(2.5,0)--++(.25,-.5)--++(.25,.5)
    --++(2.5,0)--++(.25,-.5)--++(.25,.5)--++(2.5,0)
    node[meter,anchor=east](m){}(m.west)node[below left]{$A_5$};
    \draw (0,0) --++(1,0)--++(.25,-.5)
    edge+(.25,.5)++(.25,.5)edge++(1,0)++(1,0)edge+(.25,.5)++(.25,.5)
    --++(.25,-.5)--++(1,0)--++(.25,-.5)--++(.25,.5)--++(1,0)--++(.25,.5)
    edge+(.25,-.5)++(.25,-.5)edge+(2.5,0);
    \draw (0,-1) --++(1,0)--++(.25,.5)--++(.25,-.5)--++(2.5,0)--++(.25,.5)--++(.25,-.5)--++(4,0)
    node[meter,anchor=east](m){}(m.west)node[below left]{$A_4$};
\end{tikzpicture}\bigskip

\begin{tikzpicture}[every edge/.style={draw,densely dotted}]
    \scriptsize
    \draw [ultra thick] (1,-.5)--++(.5,0)node[above right]{50:50};
    \draw [ultra thick] (2.5,.5)--++(.5,0)node[above right]{33:67};
    \draw [ultra thick] (4,-.5)--++(.5,0)node[above right]{25:75};
    \draw [ultra thick] (5.5,.5)--++(.5,0)node[above right]{33:67};
    \draw [ultra thick] (7,-.5)--++(.5,0)node[above right]{50:50};
    \draw (0,1) --++(2.5,0)--++(.25,-.5)--++(.25,.5)
    --++(2.5,0)--++(.25,-.5)--++(.25,.5)--++(2.5,0)
    node[meter,anchor=east](m){}(m.west)node[below left]{$A_5$};
    \draw (0,0) --++(1,0)--++(.25,-.5)
    edge+(.25,.5)++(.25,.5)edge++(1,0)++(1,0)edge+(.25,.5)++(.25,.5)
    --++(.25,-.5)--++(1,0)--++(.25,-.5)--++(.25,.5)--++(1,0)--++(.25,.5)
    edge+(.25,-.5)++(.25,-.5)edge++(1,0)++(1,0)edge+(.25,-.5)++(.25,-.5)
    --++(.25,.5)--++(1,0);
    \draw (0,-1) --++(1,0)--++(.25,.5)--++(.25,-.5)--++(2.5,0)--++(.25,.5)--++(.25,-.5)--++(2.5,0)
    --++(.25,.5)--++(.25,-.5)--++(1,0)
    node[meter,anchor=east](m){}(m.west)node[below left]{$A_1'$};
\end{tikzpicture}
\caption{The five different contexts for KCBS violation. The input state is $\frac1{\sqrt3}(\ket{100}+\ket{010}+\ket{001})$. For details on how to compare $A_1$ and $A_1'$ see text.}
\label{fig:KCBS}
\end{figure}

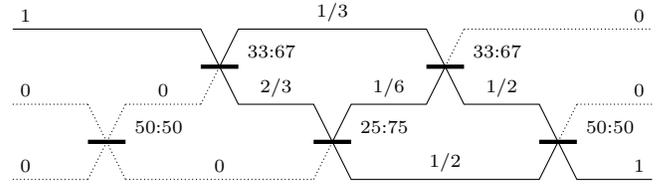
\begin{figure}
\begin{tikzpicture}[every edge/.style={draw,densely dotted}]
    \scriptsize
    \draw [ultra thick] (1,-.5)--++(.5,0)node[above right]{50:50};
    \draw [ultra thick] (2.5,.5)--++(.5,0)node[above right]{33:67};
    \draw [ultra thick] (4,-.5)--++(.5,0)node[above right]{25:75};
    \draw [ultra thick] (5.5,.5)--++(.5,0)node[above right]{33:67};
    \draw [ultra thick] (7,-.5)--++(.5,0)node[above right]{50:50};
    \draw (0,1) to["1" above right,at start]++(2.5,0)--++(.25,-.5)--++(.25,.5)
    to["1/3"]++(2.5,0)--++(.25,-.5)edge+(.25,.5)++(.25,.5) 
    edge["0" above left, at end]++(2.5,0);
    \draw (0,0)edge["0" above right,at start]+(1,0)++(1,0)
    edge+(.25,-.5)++(.25,-.5)edge+(.25,.5)++(.25,.5)edge["0"]++(1,0)++(1,0)
    edge+(.25,.5)++(.25,.5)--++(.25,-.5)
    to["2/3"]++(1,0)--++(.25,-.5)--++(.25,.5)to["1/6"]++(1,0)--++(.25,.5)
    --++(.25,-.5)to["1/2"]++(1,0)--++(.25,-.5)
    edge+(.25,.5)++(.25,.5)edge["0" above left,at end]++(1,0);
    \draw (0,-1) edge["0" above right,at start]+(1,0)++(1,0)
    edge+(.25,.5)++(.25,.5)edge+(.25,-.5)++(.25,-.5)
    edge["0"]+(2.5,0)++(2.5,0)edge+(.25,.5)++(.25,.5)--++(.25,-.5)to["1/2"]++(2.5,0)
    --++(.25,.5)--++(.25,-.5)to["1" above left,at end]++(1,0);
\end{tikzpicture}
\caption{Detection probabilities of an identical three-path interferometer with input state $\ket{100}$.}
\label{fig:single}
\end{figure}

We will use photon detectors with output $\pm1$, where $-1$ indicates ``click'' in the detector.
Three of these will be associated with the upper paths and numbered $A_1$, $A_3$ and $A_5$ from left to right, three more with the lowest paths after the first beamsplitter numbered $A_2$, $A_4$ and $A_1'$.
For a fixed $k$, the measurement $A_k$ can be performed simultaneously with $A_{k+1}$, forming a sequence of pairs $A_kA_{k+1}$ and a final pair $A_5A_1'$; so that each index $k$ is contained in two such \textit{contexts}. 

In Fig.~\ref{fig:KCBS} the detectors $A_k$ are unaffected by the context change in the sequence, so it seems well-motivated for us to use a noncontextual model for the corresponding measurements. 
The first and final measurements $A_1$ and $A_1'$ are implemented differently, but a single photon that enters the topmost input port of the interferometer will exit through the bottom output port, since
\begin{equation}
\begin{split}
\ket{100}&\xrightarrow{BS_1}\ket{100}\\
&\xrightarrow{BS_2}\sqrt{\tfrac23}\ket{010}+\sqrt{\tfrac13}\ket{100}\\
&\xrightarrow{BS_3}\sqrt{\tfrac12}\ket{001}+\sqrt{\tfrac16}\ket{010}+\sqrt{\tfrac13}\ket{100}\\
&\xrightarrow{BS_4}\sqrt{\tfrac12}\ket{001}+\sqrt{\tfrac12}\ket{010}\\
&\xrightarrow{BS_5}\ket{001}
\end{split}
\end{equation}
giving the path probabilities in Fig.~\ref{fig:single}. 
Thus, in a noncontextual model for the ideal setup it would be natural to assume that $A_1=A_1'$ (with probability~1).

For a more robust way to handle differences between different contexts in situations like this, that is also symmetric between all $A_k$, one can use the concept of \textit{maximal noncontextuality} \cite{Kujala2015}.
Especially when performing an actual experiment, where none of the marginals $E(A_k)$ typically coincide between different contexts (e.g., \cite{Lapkiewicz2011}), a noncontextual model may seem less motivated, but the more robust concept of maximal noncontextuality resolves this issue. 
Under the assumption of maximal noncontextuality, the theoretical prediction (equality with probability 1) is sufficient to motivate the assumption that $A_1=A_1'$.
Lapkiewicz et al~\cite{Lapkiewicz2011} performs an experiment for a setup which is in essence equivalent. However, our use here of Hofmann's version \cite{Hofmann2023Sequential} shows more clearly that we would expect that $A_1=A_1'$.

For a noncontextual model the KCBS inequality applies,
\begin{equation}
\begin{split}\label{eq.KCBS}
&E(A_1A_2)+E(A_3A_2)\\&+E(A_3A_4)+E(A_5A_4)+E(A_5A_1)\ge-3    
\end{split}
\end{equation}
but from the above discussion we arrive at
\begin{equation}
\begin{split}
  &(-\tfrac13+0-\tfrac23)+(-\tfrac19+\tfrac29-\tfrac23)+(-\tfrac19+\tfrac29-\tfrac23)\\
  &+(-\tfrac13+0-\tfrac23)+(-\tfrac13+\tfrac13-\tfrac13)
  =-3\tfrac49\le-3    
\end{split}
\end{equation}
This is a clear violation of the KCBS inequality \cite{Klyachko2008KCBS}, meaning the scenario is contextual (the maximal quantum violation is $5-4\sqrt5\approx-3.944$).
The violation shows that no noncontextual hidden-variable model can give the probabilities indicated in Fig.~\ref{fig:interferometer}.
More simply put, a model which assigns a definite path to a photon will sometimes need to reassign path (the ``value'' of $A_k$) depending on whether the measurement context is $A_{k-1}$ or $A_{k+1}$.
The photons will need to ``jump around'' in the interferometer.

Given that we can build a single-particle interferometer which violates a noncontextuality inequality, this shows not only can we build an interferometer which cannot be reproduced by Spekkens' toy model (given the toy model can only reproduce noncontextual scenarios); but also that such an interferometer demonstrates aspects of the phenomenology of interference which cannot be described classically.

\section{Discussion}


A key motivation for this work was to investigate what the aspects of the phenomenology of interference are, that are required for such a scenario to be contextual, here taken to imply nonclassical.

The previous works which claimed to show that some, but not all, of the phenomenology of interference could be reproduced classically (\cite{Catani2023interference} and \cite{Catani2023InterNonclassical} respectively) were both relatively vague about what the actual aspects of interference were that enabled/blocked the phenomenon from being reproduced classically. One specific thing which Ref.~\cite{Catani2023InterNonclassical} showed was that certain interference phenomena cannot be reproduced by a model which is Spekkens' noncontextual. Specifically, Catani \emph{et al} show that the wave-particle duality relation, which, unlike Greenberger and Yassin \cite{Greenberger1988Simultaneous}, they interpret as an uncertainty relation, cannot be reproduced by a Spekkens'-noncontextual model. 
We would argue that violations of Spekkens' noncontextuality do not provide as strong a signature for the inability to represent a given scenario classically as Kochen-Specker contextuality. 
This is as violating Spekkens' noncontextuality just requires the inability to make a hidden variable model with always-positive probabilities across all possible preparations and measurements for a scenario, so one negative quasiprobability is enough to violate it, even if that quasiprobability being negative doesn't correspond to any contradiction in predictions of outcomes.
Violating Kochen-Specker contextuality on the other hand requires an actual contradiction in predictions of experimental outcomes when we compare between contexts, as above. 
We therefore would suggest that the argument by Catani \emph{et al} that certain aspects of the phenomenology of interference cannot be reproduced classically is less compelling than the one we give here.

One interesting observation is that the interferometer we use is a simple example of a Boson sampling circuit~\cite{brod2019photonic}. 
In Boson sampling one uses an interferometric setup that typically contains many layers of beamsplitters with different transmission ratios, send number states into the interferometer, and measure photon presence at the output ports. 
The simple Mach-Zehnder interferometer used in the extended Elitzur-Vaidman Bomb-tester can be viewed as the simplest possible example, with two beamsplitters in two layers; the present paper shows that such a setup is classically simulable, meaning with a Kochen-Specker noncontextual model.
Hofmann's three-path interferometer could be seen as the first nontrivial example that resists classical, or Kochen-Specker-noncontextual, simulation.
When used in a Boson sampling setup, the state in Eq.~\eqref{eq:contextual_state} used to show contextuality of the setup needs to be generated, which can be done from a single-particle product state using two beam-splitters, so that the whole setup has three paths and seven beamsplitters arranged in seven layers.
The Kochen-Specker contextuality of this relatively simple Boson sampling problem tells us that Kochen-Specker contextuality will be present in many such arrangements, for the first time establishing a link between Boson sampling and Kochen-Specker contextuality.
An interesting question is if Kochen-Specker contextuality is necessary for the advantage of Boson sampling. This of course warrants further study.

In conclusion, we above reproduced the phenomenology of the Elitzur-Vaidman Bomb Tester (including unbalanced versions) using a relatively straightforward extension of the quantum simulation logic (QSL) formalism. This showed that some of the phenomenology of interference can be reproduced in a ``classical'' way. Our result improved upon and simplified a previous result \cite{Catani2023interference}, which relies on a much more complicated extension on Spekkens' Toy Model referred to as a ``toy field theory.'' We also showed that not all single-particle interference can be explained by such a simple extension, by showing that Hofmann's three-path interferometer is ``nonclassical'' in a very specific sense: it violates a Kochen-Specker-noncontextual inequality. Given that both our extension of QSL and Catani et al's extension are Kochen-Specker \emph{noncontextual} --- so do not reproduce the contextual behaviour of Hofmann's three-path interferometer --- the behaviour of that interferometer is a proper example of a phenomenon which Feynman considered to have in it the heart of quantum mechanics.

\textit{Acknowledgements:} JRH acknowledges support from a Royal Society Research Grant (RG/R1/251590), and from their EPSRC Quantum Technologies Career Acceleration Fellowship (UKRI1217). JK and J\AA L acknowledge support from the Swedish Science Council project no 2023-05031.

\bibliographystyle{unsrturl}
\bibliography{ref.bib}

\end{document}